# Discharge characteristics of a low-pressure geometrically asymmetric cylindrical capacitively coupled plasma with an axisymmetric magnetic field

Swati Dahiya[1,2], Pawandeep Singh[1,2], Yashashri Patil[1], Sarveshwar Sharma[1,2], Nishant Sirse[3] and Shantanu Kumar Karkari[1,2]

[1]Institute for Plasma Research, Bhat, Gandhinagar, Gujarat, 382428 India

[2]Homi Bhabha National Institute, Training School Complex, Anushaktinagar, Mumbai-400094, India

[3]Institute of Science and Research and Centre for Scientific and Applied Research, IPS Academy, Indore-452012, India

## Abstract

We investigate the discharge characteristics of a low-pressure geometrically asymmetric cylindrical capacitively coupled plasma discharge with an axisymmetric magnetic field generating an $E \times B$ drift in the azimuthal direction. Vital discharge parameters, including electron density, electron temperature, *DC* self-bias, and Electron Energy distribution function (*EEDF*), are studied experimentally for varying magnetic field strength (*B*). A transition in the discharge asymmetry is observed along with a range of magnetic fields where the discharge is highly efficient with lower electron temperature. Outside this range of magnetic field, the plasma density drops, followed by an increase in the electron temperature. The observed behaviour is attributed to the transition from geometrical asymmetry to magnetic field-associated symmetry due to reduced radial losses and plasma confinement in the peripheral region. In this region, the *DC* self-bias increases almost linearly from a large negative value to nearly zero, i.e., the discharge becomes symmetric. The *EEDF* undergoes a transition from bi-Maxwellian for unmagnetized to Maxwellian at intermediate *B* and finally becomes a weakly bi-Maxwellian at higher values of *B*. The above transitions present a novel way to independently control the ion energy and ion flux in a cylindrical *CCP* system using an axisymmetric magnetic field with an enhanced plasma density and lower electron temperature operation that is beneficial for plasma processing applications.

## I. Introduction

As the semiconductor industry has evolved over the last few decades, low-temperature, low-pressure capacitively coupled plasma (*CCP*) devices have been introduced and are widely employed for a variety of applications, such as etching and deposition processes [1-6]. The research has concentrated on optimizing plasma parameters such as plasma density, ion flux, and ion energy to improve processing productivity and independent control over these parameters [7-14]. Higher input power supports higher density in such devices, which may damage the surface to be treated.

To enhance the efficiency, maintain high density at lower *RF* powers, and independent control of energy and flux of ions, several improvisations in *CCP* discharges have been done over time, such as using multiple *RF* frequencies [15-26], tailored voltage and current waveforms [27-37] and pulsed *RF* discharges [38-43]. The use of higher driving frequency in the very high frequency (*VHF*) range produces higher plasma density, and therefore the ion flux towards the substrate increase, which enhances the etching rates [13, 44-53]. However, *CCPs* excitation in the *VHF* range may lead to plasma non-uniformity due to standing wave and skin effects [54, 55]. Shaped and segmented type powered electrodes are proposed to overcome the non-uniformity issues in the *VHF CCP* discharges [56-60]. Another possibility to enhance and control the plasma parameters could be using a static magnetic field perpendicular to the sheath electric field. Low magnetic fields of a few milli Tesla (mT) are utilized to improve the plasma density inside such devices by reducing the loss of electrons to the electrodes [61, 62].

Due to the broad range of applications, several studies have been conducted on the influence of magnetic fields in *CCP* discharges. Liebermann suggested in 1991, using the homogeneous sheath model, that applying an external magnetic field parallel to the discharge plates improves collision-less heating [63]. However, through simulation in 1995 and 1996 *Turner* demonstrated that a small amount of transverse magnetic field can induce a heating mode change from pressure-heating to collisional heating [9,64], further explored experimentally in a parallel plate configuration [65]. Kushner used 2D-hybrid fluid modeling to investigate the effect of an externally applied magnetic field on the ionization processes inside magnetically enhanced *CCP* sources [66]. You et al. [67-70] and Park et al. [71] conducted simulation and experimental studies in parallel plate *CCP* discharges for various pressure regimes. It is established that the effect of the B-field is equivalent to an increase in neutral gas pressure [69], and applying an external magnetic field induces a non-local to local transition of kinetic properties [67]. Barnat et al. investigated the effect of a magnetic field in a *GEC* reference cell [72]. Their study was conducted using several diagnostic tools, and the results demonstrated that the

*RF* voltage decreases with the applied B-field, as does the phase difference between the *RF* voltage and current waveforms. A drop in the phase between current and voltage indicates that the discharge becomes more resistive., Zheng et al. demonstrated the significance of metastable atoms and multistep ionization [73] at high B-field and high pressure. In another study, a magnetic field is utilized to attain Electron Series Resonance (*ESR*), illustrating that as the externally applied magnetic field increases, the phase between the applied rf voltage and current decreases, and the discharge becomes purely resistive, resulting in maximum power transfer to the discharge [74].

The parallel plate configuration is the most widely used electrode configuration for *CCP* discharges, with one electrode grounded and the other powered by the *RF*. Sharma et al. [75] used particle-in-cell simulation technique and demonstrated that the effective control of ion energy and ion flux enhancement can be achieved by applying a transverse external magnetic field in a parallel plate *CCP* system. On applying a high magnetic field parallel to the electrode, the $E \times B$ forces generate plasma non-uniformities on the electrode surface [72, 76, 77]. In some discharges like Magnetically Enhanced Reactive Ion Etcher (*MERIE*), rotating static magnetic fields remove these non-uniformities [78]. A cylindrical *CCP* configuration with a magnetic field directed in the z-direction is another way to eliminate lateral inhomogeneities [73, 79]. In such devices, the $E \times B$ forces will be in the azimuthal direction, and the closed motion will maintain overall homogeneity over the cylindrical surface and further enhance the ionization probability/discharge efficiency. Recently it has been reported that operating *CCP* with *VHF* in the presence of a transverse uniform weak magnetic field triggers a new operational regime where a significant enhancement in the performance of low-pressure *CCP* discharge has been observed [80-82]. Patil et al. [80] were the first to report that performance is enhanced due to a resonance effect called Electron Cyclotron Bounce Resonance (*ECBR*), which appears when electron cyclotron frequency ($f_{ce}=eB/m_e$) matches half of the *RF* frequency. Opposite to *MERIE* configuration, the applied external magnetic field is very weak (~10 G), so the generation of non-uniformity in plasma is significantly small here [80, 82].

In a recent publication, the electron heating near to the peripheral region in a cylindrical *CCP* system with no radial confinement was investigated experimentally [83]. In the present manuscript, we report the experimental study of bulk plasma parameters, including Electron Energy Distribution Function (*EEDF*) and *DC* self-bias in a magnetized cylindrical *CCP* discharge system with radial confinement. The magnetic field is applied perpendicular to the sheath electric field, i.e., in the axial direction. The study is performed using an *RF*-compensated Langmuir probe, commercial Inline IV sensor, and high voltage probe for several *RF* powers from 20 W to 100 W and from an unmagnetized case to a range of magnetic field strength up to 11 mT.

## II. Plasma discharge system and diagnostic techniques

a) <u>Plasma discharge system</u>

Figures 1 (a) and 1 (b) show the schematic of the plasma discharge system and discharge circuit, respectively. As shown in Figure 1 (a), the experimental set-up consists of a Cylindrical Powered Electrode (*CPE*), which is 20 cm long with an inner diameter of 24 cm. The thickness of the electrode is 3 mm. Two annular rings with inner diameters of 19 cm and outer diameters of 23 cm are attached at the end of *CPE*. 2 sets of grids (Grid 1 (G1) and Grid 2 (G2)) are installed after the annular rings for the plasma confinement. G1 is plasma facing fine grid connected to the ground, whereas G2 is the coarse grid that is kept floating. The CPE, annular rings, G1 and G2, are kept electrically isolated using an insulator made of PTFE. The whole assembly is kept inside a bigger cylindrical vacuum chamber having an inner diameter of 31 cm and a length of 120 cm. The electrode assembly is kept isolated from the vacuum chamber using ceramic (Alumina $Al_2O_3$) beads. The chamber is pumped up to base pressure of $5 \times 10^{-4}$ mTorr through a set of 700 l/s Pfeiffer turbo molecular pump (*TMP*) (Turbo Hi-pace 700) and rotary pump. The discharge is performed in argon gas, flowing into the chamber via one of the axial ports located on the other side of the *TMP*.

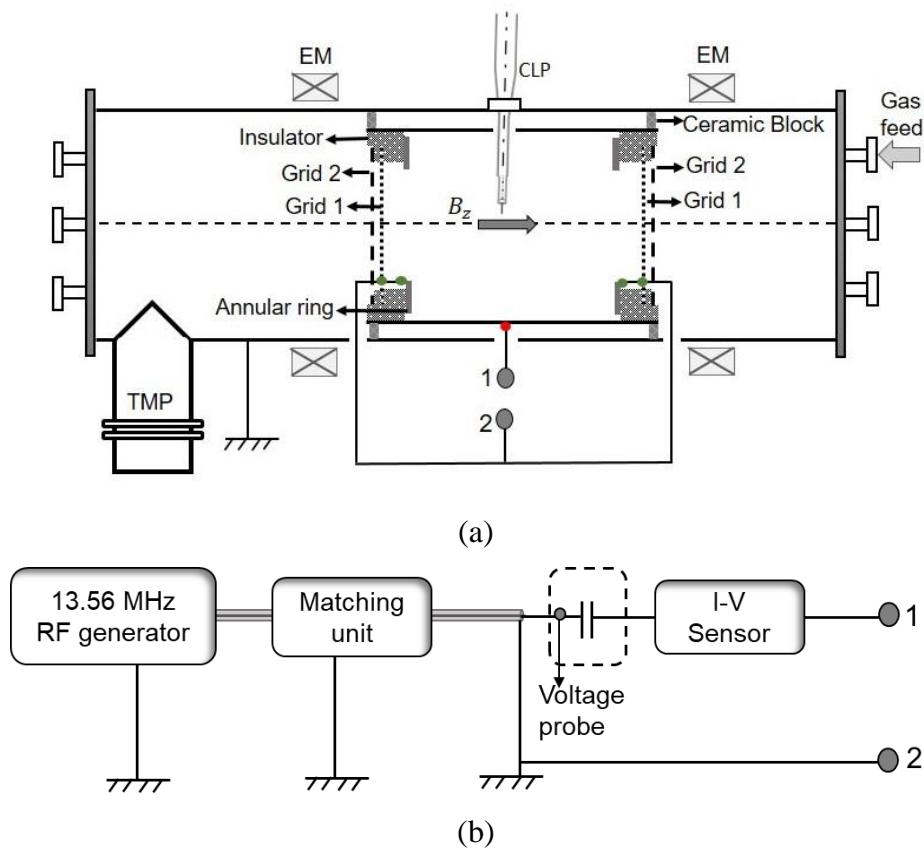

Figure 1: (a) Experimental setup where *EM* denotes the electromagnetic coils, *TMP* is turbo molecular pump, and *CLP* is the compensated Langmuir probe; (b) the discharge circuit.

An axially uniform axial magnetic field is obtained using a set of electromagnetic coils arranged in Helmholtz configuration having a 27 cm distance between them. The coils are made by winding copper bus bars in a double pan-cake configuration. These coils can produce a 0.26 mT magnetic field strength for one ampere of current. The coil current is provided by using a regulated power supply (Aplab L3260) which can supply a maximum current of 60 A, which corresponds to a maximum of 15.6 mT magnetic field. The magnetic field profile inside the vacuum chamber is shown in Figure 2. The electrode assembly is kept in the uniform magnetic field region, i.e., the shaded region shown in Figure 2.

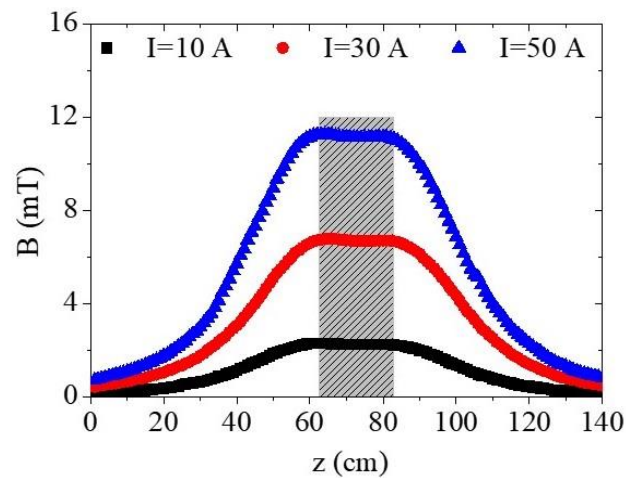

Figure 2: Plot showing axial magnetic field profile in the vacuum chamber at different coil current values. The shaded region shows the uniform $B_z$ region where the electrode assembly is placed.

For characterizing the discharge, a small cavity is kept at the center of the cylindrical electrode surface to insert the Langmuir probe radially, as shown in the Figure 1 (a). Argon discharge is created using a 13.56 MHz RF generator (Coaxial Power systems-AG1213W) coupled to the electrodes through a pi-type matching network. The cylindrical electrode is powered via a blocking capacitor of 3 nF with respect to the grounded annular rings and G1.

b) Diagnostic techniques

The *RF* voltage (V), current (I), and phase between them are measured using a commercial inline I-V sensor (Octiv poly 2.0, Impedans Ltd Ireland). As shown in figure 1 (b), the inline I-V sensor is installed between the matching network output and plasma load to measure the actual power deposited into the discharge. A high-voltage probe (TEK P5205A) is used to measure the *DC* self-bias generated on the powered electrode with respect to the grounded chamber.

An *RF*-compensated Langmuir probe with a main probe and a floating auxiliary probe is assembled with a resonant filtering technique to compensate for the effect of *RF* frequency (13.56 MHz) and its second harmonic (27.12 MHz) on the *IV* characteristics. The main probe is made of tungsten cylindrical wire with a diameter of 0.2 mm and a length of 5 mm. The auxiliary probe is a 0.2 mm diameter tungsten wire coupled to the measurement probe by a small 1nF capacitor. The purpose of auxiliary probe is to lower the input *RF* impedance across the probe sheath. A resonant circuit comprising four inductors is connected in series near the probe tip. The compensating probe has an impedance of about 110 k Ω at 13.56 MHz, which satisfy the relation $Z_f \geq \frac{Z_{pr}\, e\, V_{pl\_rf}}{T_e}$ ; where $V_{(pl\_rf)}$ is the *RF* plasma potential reference to ground, $Z_{pr}$ is the impedance between the probe and plasma and $Z_f$ is the probe impedance. The *IV* characteristics are obtained by biasing the probe with respect to the grounded chamber. The plasma density is calculated by applying the orbital motion limited theory to the ion saturation region of the *IV*-characteristic [84, 85]. The electron temperature is calculated from the semi-log plot of the electron current, and the plasma potential is determined from the double derivative of the *IV* curve [86]. The *EEDF* is measured by the method described in [83].

## III. Results and Discussions

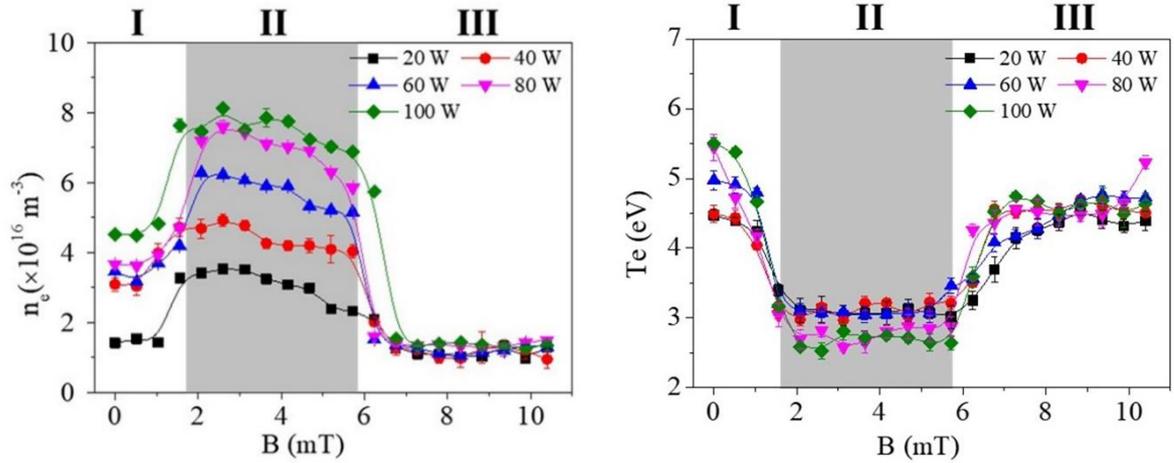

Figure 3: Plot showing (a) plasma density and (b) electron temperature at the center (*r* = 0, *z* = 0) of the discharge versus magnetic field for different RF powers. The operating gas pressure is 1 Pa.

Figures 3 (a) and 3 (b) show variations in the electron density and temperature respectively at the center of the discharge versus magnetic field strength for different values of discharge power. The magnetic field strength is increased from 0 to 11 mT in the step of 0.5 mT, and the *RF* power is varied from 20 W to 100 W in the step of 20 W. At all magnetic field strengths, the *RF* power is monitored separately using the *IV* sensor after the matching network and kept constant by adjusting the output power from the *RF* generator. The operating gas pressure is kept constant at 4.0 mTorr. The experimental results show a non-monotonic trend of the plasma density and electron temperature versus magnetic field strength for all RF powers. More specifically, there is a range (~1.6 to 5.9 mT) of magnetic is observed for which the density stays higher (~two-fold in comparison to non-magnetized/weakly magnetized case) and nearly constant (prominent at 100 W). With a further rise in the magnetic field strength after this saturation, the electron density drops significantly, even lower when compared to non-magnetized/weakly magnetized case. The electron temperature (Figure 3 (b)), on the other hand, shows an opposite trend, i.e., it stays lower for the range of the magnetic field when the electron density is higher and vice versa. The maximum plasma density is observed for 100 W RF power, which is ~$8\times10^{16}$ m$^{-3}$, whereas the electron temperature for the same operating conditions is lowest (~2.5 eV). Thus, a specific range of the magnetic is observed for which the plasma density remains highest while maintaining lower values of electron temperature.

An opposite trend of the plasma density and electron temperature is associated with the different power consumption in the ionization and heating process and can be explained from the global power balance model [1], which is as follows –

$$n = \frac{P_{abs}}{e\, u_B A_{eff} \varepsilon_T}$$

where $n$ is the plasma density, $P_{abs}$ is the power absorbed in the discharge, $u_B$ ($\sqrt{kT_e/M}$, $kT_e$ is electron temperature in electron volts and $M$ is the mass of the ion) is the Bohm speed, $A_{eff}$ is the effective area for the particle loss at the electrodes and $\varepsilon_T$ is the sum of collisional energy loss per electron-ion created ($\varepsilon_c$), mean kinetic energy lost per ions ($\varepsilon_i$) and electrons ($\varepsilon_e$). In the above relation, the plasma density and electron temperature are inversely proportional for a constant discharge power and effective area. In this experiment, the RF power transferred to the discharge is kept constant for varying magnetic fields; hence the electron temperature decreases with a rise in plasma density and vice versa.

The effect of the magnetic field on the plasma density and electron temperature can be explained by dividing the magnetic field into three regions. In region I, for unmagnetized/weakly magnetized case (up to 1.6 mT), the plasma density stays lower for different values of RF power. It means the discharge is not very efficient due to various losses mechanism, including wall losses. In this region, the plasma density increases with an increase in the RF power level. It is observed that the plasma density rises from ~$1.5\times10^{16}$ m$^{-3}$ at 20 W to ~$4.5\times10^{16}$ m$^{-3}$ at 100 W RF power, i.e., a three-fold increase in the plasma density is observed. In region II, electrons are magnetized while the ions are unmagnetized as $r_{Li} > R_{CPE}$

($R_{CPE}$ is the radius of *CPE*), the radial losses of the ionizing electrons reduce, and the bulk density increases up to twice when compared to region I. For a range of applied magnetic fields, the density saturates or decreases slightly. As the electron mobility in the sheath is reduced, resulting in suppression of the energy transfer from the sheath to the plasma bulk. Therefore, the plasma density does not increase further in this region. In region III, the density drops with a further rise in the magnetic field strength. This is due to the fact that electrons are strongly magnetized while ions are very weakly magnetized, and the strong magnetic field restricts the radial flow of high-energy electrons responsible for the ionization process, which reduces the bulk plasma density. As shown in figure 4, at higher magnetic fields, the Larmor radius of the ionizing electrons becomes lower than the sheath width ($r_{Le} < r_{sh}$). In this plot, the sheath width is calculated using the Child-Langmuir formula [1], i.e., $r_{sh} = \frac{\sqrt{2}}{3} \lambda_{Ds} \left(\frac{2V_0}{T_e}\right)^{3/4}$, $\lambda_{Ds}$ being the Debye length $\lambda_{Ds} = \left(\frac{\varepsilon_0 T_e}{e n_e}\right)^{1/2}$; where $V_0$ is the electrode voltage, $n_e$ is plasma density and $T_e$ is electron temperature, these are the experimentally measured values. Hence, all the high-energy electrons are confined in the peripheral region. Also, the electron momentum transfer collision frequency falls below the cyclotron frequency ($\nu_m \ll f_{ce}$), such that the effective ionizing collisions decrease, as also observed in [66]. Hence, it can be stated that above these threshold magnetic field strengths, the ionization probability decreases very sharply; therefore, a significant drop in the plasma density is observed.

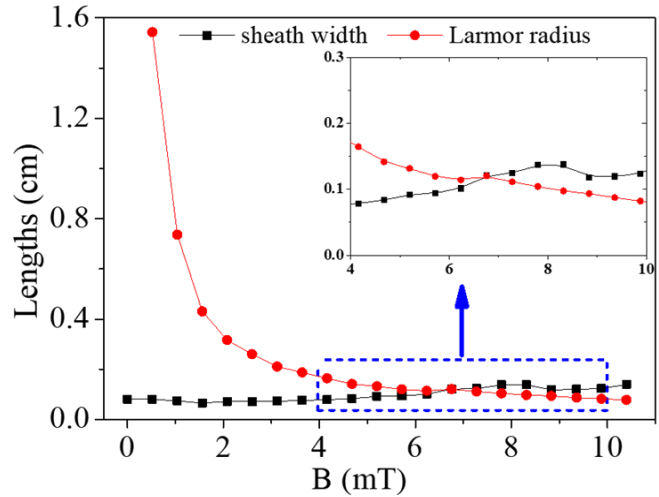

Figure 4: Plot showing the Larmor radius and sheath width for different *RF* powers with increasing magnetic field strength. The inset graph shows the magnetic field, after which $r_{Le}$ becomes lower than $r_{sh}$.

To understand the details of electron energy deposition among different energy range, diagnostics of Electron Energy Probability Function (*EEPF*) is performed for two radial positions. Figures 5 (a) and 5 (b) show the *EEPF* at the center of the discharge ($r = 0$ cm) and near the edge ($r = 8$ cm), respectively. The measurements are performed for an unmagnetized case (region I), $B = 3$ mT (region II), and $B = 8$ mT (region III). Regions I, II and III are labeled in figure 3 and correspond to different plasma density/temperature regimes. The data presented in Figure 5 are smoothened by the *Savitzky-Golay* method with 200 points on window and second-order polynomial. As shown in Figure 5 (a), for the unmagnetized case ($B = 0$ mT), the *EEPF* depicts bi-Maxwellian distribution. A bi-Maxwellian shape of the *EEPF* in an unmagnetized *CCP* discharge is attributed to the ambipolar electric field that confines the low-energy electron, whereas high energy electrons could overcome the ambipolar field and gain energy from the oscillating sheath resulting in 2 temperature Maxwellian. In region II ($B = 3$ mT), where the plasma density is higher (Figure 3 (a)), the population of low energy electrons increases at a much faster rate in comparison to the high energy electrons, and therefore the *EEPF* turns in to a Maxwellian distribution. In this region, the population of high-energy electrons is lower when compared to the unmagnetized case. As discussed earlier, due to the confinement of electrons by the magnetic field, the wall losses decrease, and ionization probability increases in bulk, increasing the population of low-energy electrons. In region III, the population of low energy electrons decreases drastically; the *EEPF* turns into nearly bi-Maxwellian. It is because the higher magnetic field (lower electron gyro radius) confines high-energy electrons near the peripheral region, decreasing the bulk ionization.

Near the edge ($r = 8$ cm), Figure 5 (b), the population of low-energy electrons decreases for all three regions; however, a significant different is observed in the high-energy tail. The tail-end electrons population is highest in the case of $B = 3$ mT when compared to $B = 0$ mT and 8 mT. Interestingly it is also higher when compared to $r = 0$ cm. It confirms that the electrons are strongly heated near the edge, getting confined and generating higher plasma density in region II. For region III ($B = 8$ mT), the *EEPF* depicts a similar shape as the $r = 0$ cm case with a lower population of low and high-energy electrons. It is worth noticing that the *EEDF* is measured 4 cm away from the *CPE* surface, and a signature of the lower population of high-energy electrons confirms that the high-energy electrons are confined in the peripheral region due to higher magnetic field strength ($B = 8$ mT)/ smaller electron Larmor radius.

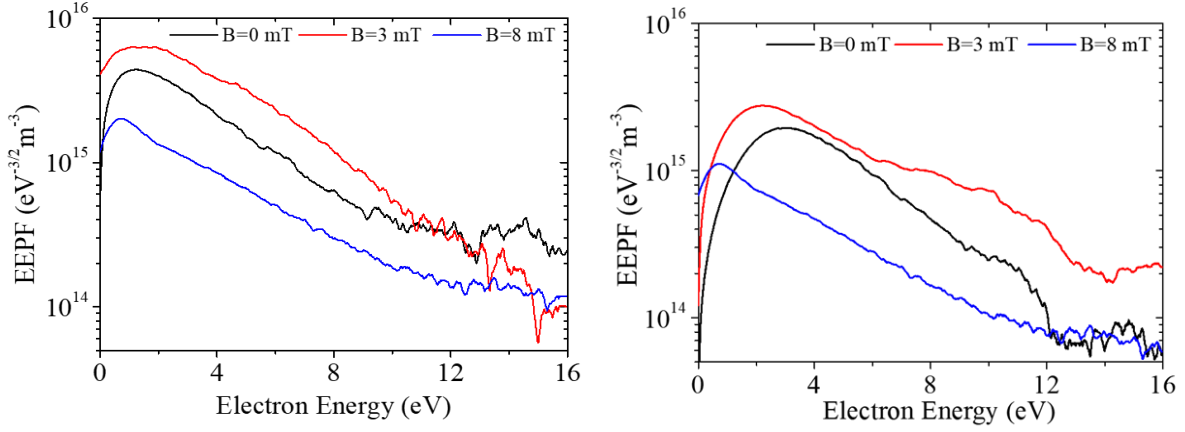

Figure 5: Plot showing Electron Energy Probability Function (*EEPF*) for different values of magnetic field strength at (a) *r* = 0 cm (center) and (b) *r* = 8 cm (near the edge).

The plasma confinement and associated change in the bulk plasma parameters with the magnetic field affects the charged particle losses on the wall, and therefore, the *DC* bias potential on the powered electrode varies accordingly. Figure 6 shows a variation in the *DC* self-bias on the powered electrode for different values of magnetic field strengths from 0 to 11 mT and *RF* powers from 20 W to 100 W. The shaded part is region II where the plasma density is highest (Figure 3). The *DC* self-bias is negative in the absence of a magnetic field. This is due to the fact that the area of powered ($A_p$) and grounded ($A_g$) electrodes are different, i.e., $A_p \approx 3A_g$; therefore, the geometric asymmetry is present in the discharge. The *DC* self-bias increases from -10 V at 20 W to -55 V at 100 W *RF* power level. With the increase in the magnetic field strength, the *DC* self-bias decreases slightly; however, in region II, it changes linearly from a large negative value to nearly zero, i.e., the discharge becomes symmetric. With a further rise in the magnetic field strength (region III); no change in the *DC* self-bias is observed. This effect is attributed to the electron confinement by the magnetic field. As the magnetic field is applied axially, it restricts the radial motion of electrons towards the cylindrical powered electrode surface. However, the motion of electrons toward the grounded electrode, i.e., on the axial ends, remains unaffected. Hence, the *DC* self-bias reduces, and the discharge tends to be more symmetric. From a processing perspective, it is highly useful as the plasma density in region II remains higher and nearly constant, which enhances the flux, and the *DC* self-bias could be varied by changing the magnetic field that controls the ion energy.

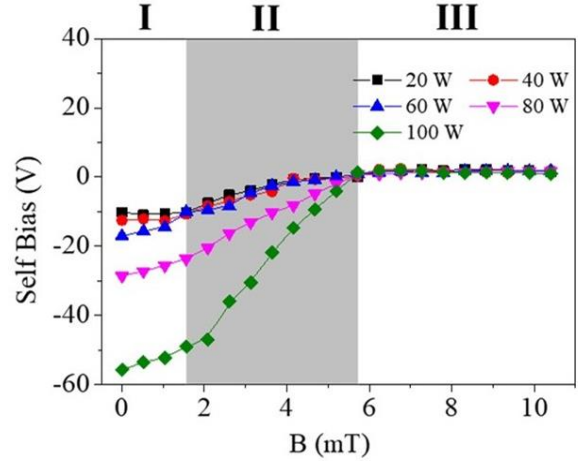

Figure 6: Plot showing the DC self-bias generated at the powered electrode versus magnetic field for different RF powers.

## IV. Summary and Conclusions

The discharge characteristics of a low-pressure cylindrical *CCP* system are investigated experimentally in the presence of an axisymmetric magnetic field. The plasma density and electron temperature measurements using an *RF*-compensated Langmuir probe shows a transition and an enhanced operating regime for a range of magnetic field strengths. In this range, the plasma density remains higher (almost double) and nearly constant when compared to unmagnetized/weakly magnetized and higher magnetic field strength cases. On the other hand, the electron temperature shows an opposite trend, i.e., it remains lowest in the constant plasma density region and vice versa. The observed phenomenon is attributed to a balance between geometrical asymmetry and the magnetic field-associated reduction in the wall losses of the charged particles. Furthermore, confinement in a closed $E \times B$ drift nearly to the peripheral region allows enhanced ionization rates suitable for high-density plasmas. The *EEDF* shape undergoes a transition from bi-Maxwellian in the unmagnetized case to a Maxwellian type in the high-density region. Finally, it becomes a weakly bi-Maxwellian at higher magnetic field strengths. The *EEDF* measurement near the plasma periphery depicts a drastic increase in the population of high-energy electrons, which supports the higher ionization, and plasma density enhancement in the intermediate magnetic field region. In conclusion, the experiments provide a range of magnetic fields to attain high-density and low-temperature plasma in cylindrical *CCP* systems that is beneficial for many applications where a high ion flux along with low energy electrons is required.

## Acknowledgment

This work is supported by the Department of Atomic Energy, Government of India, and Science and Engineering Research Board (SERB) Core Research Grant No. CRG/2021/003536.